\documentclass[12pt,
aps,preprint
showpacs,nofootinbib,floatfix]{revtex4}

\usepackage{graphicx}
\usepackage{float}
\usepackage{amsmath}
\usepackage{epsfig}

\newcommand{\be}{\begin{equation}}
\newcommand{\ee}{\end{equation}}
\newcommand{\ba}{\begin{eqnarray}}
\newcommand{\ea}{\end{eqnarray}}

\newcommand{\tbar}{\bar{t}}
\newcommand{\qbar}{\bar{q}}
\newcommand{\rarr}{\rightarrow}
\newcommand{\beq}{\begin{eqnarray}}
\newcommand{\eeq}{\end{eqnarray}}
\newcommand{\rL}{\mathrm{L}}
\newcommand{\Nf}{N_\mathrm{f}}

\newcommand{\gs}{g_\mathrm{s}}
\newcommand{\calA}{\mathcal{A}}
\newcommand{\calB}{\mathcal{B}}
\newcommand{\calD}{\mathcal{D}}
\newcommand{\tree}{\mathrm{tree}}
\newcommand{\virt}{\mathrm{virt}}
\newcommand{\ri}{\mathrm{i}}
\newcommand{\rd}{\mathrm{d}}

\newcommand{\Nc}{N_\mathrm{c}}
\newcommand{\nl}{\nonumber \\}
\newcommand{\refeq}[1]{Eq.(\ref{#1})}
\newcommand{\refeqs}[2]{Eqs.(\ref{#1})-(\ref{#2})}

\newcommand{\qparbar}{\hspace{0.3em} q \hspace{-0.7em} {^{^{(-)}}}}
\newcommand{\tparbar}{\hspace{0.3em} t \hspace{-0.7em} ^{^{^{(-)}}}}
\newcommand{\bparbar}{\hspace{0.4em} b \hspace{-0.7em} ^{^{^{(-)}}}}
\newcommand{\nuparbar}{\hspace{0.4em} \nu \hspace{-0.7em} ^{^{(-)}}}
\newcommand{\Fslash}[1]{#1 \hspace{-0.42em}/\hspace{-0.08em}}

\begin{document}
\begin{titlepage}

\begin{flushright}
\vbox{
\begin{tabular}{l}
\end{tabular}
}
\end{flushright}

\vspace{0.6cm}

\title{NLO QCD corrections to top quark pair production and decay at hadron colliders
}

\author{Kirill Melnikov\footnote[4]{melnikov@pha.jhu.edu}
and Markus Schulze\footnote[3]{schulze@pha.jhu.edu} }
\affiliation{Department of Physics and Astronomy,\\
Johns Hopkins University,\\
Baltimore, MD 21218, USA}

\begin{abstract}
\vspace{2mm}
We present results for the next-to-leading order QCD corrections
to the production and semi-leptonic decays of a top quark pair
in hadron collisions, retaining all spin correlations.
To evaluate the virtual corrections, we  employ
generalized $D$-dimensional unitarity.
The computation is implemented in a numerical program
which allows detailed studies of $t \bar t $-related
observables at the Tevatron  and the LHC.

\end{abstract}

\maketitle

\thispagestyle{empty}
\end{titlepage}

\section{Introduction}
The production of top quark pairs is an interesting
process for understanding QCD dynamics. The very short life-time
of a top quark and its large mass enable
accurate theoretical predictions for this process for all energies,
including close to $t \bar t$ threshold. One can then use the
$t \bar t$ production to study properties of the
top quark, such as the value of the top quark mass,
branching fraction   $\Gamma(t \to W b)/\Gamma_{\rm tot}$ etc.
In addition, the short life-time and large mass of the top quark have important
consequences for top quark polarization since non-perturbative fluctuations
of chromomagnetic field are too weak to change  the direction of the
top quark spin.
Hence, in the absence of hard gluon radiation,
top quark polarization is conserved; information about
it can be obtained from properties of the top quark
decay products. This information
can be used to study the Lorentz structure of interaction vertices
involved in top quark production and decay. Also, one can use top quark spin
correlations to distinguish  different production mechanisms,
e.g. $q \bar q \to t \bar t$  from $g g \to t \bar t$, at least
in certain kinematic limits.  This rich physics is fully reflected
in an experimental top quark program at the Tevatron,  which is
the primary source of our knowledge about top quark properties.
The summary of recent experimental results  can be found in
Refs.~\cite{Shary:2009zj,Wagner:2005jh,Kehoe:2007px}.

On the other hand, as was repeatedly emphasized in the past, top quark physics
at the LHC
is not just a rescaled version of the top quark physics at the
Tevatron. Indeed,
the high  energy and luminosity of the LHC
will lead to a dramatic increase in   the number of
observed $t \bar t$ events.
As the result, interpretations
of cross-section measurements
will be subject to theoretical uncertainties rather than statistical errors.
This fact motivated recent analysis \cite{Moch:2008qy,Cacciari:2008zb,Kidonakis:2008mu} and reviews
\cite{Wagner:2005jh,Kehoe:2007px,bernr}
where the quality of current
understanding of $t \bar t$ production was thoroughly assessed.

We point out in this regard that pioneering studies of heavy quark
production cross-sections at next-to-leading order (NLO) in QCD were performed
almost twenty years ago \cite{Nason:1989zy,Beenakker:1990maa}\footnote{It is interesting to point
out that {\it analytic} results
for NLO QCD corrections
to top production in $gg, q\bar q$ and $q g$ channels were obtained  very
recently in Ref.\cite{Czakon:2008ii}.}.  These computations were
further extended to cover various kinematic distributions of the produced
top quarks in Refs.~\cite{Mangano:1991jk,Frixione:1995fj}. In addition, threshold resummation
was applied to top quark production at the Tevatron and the LHC
\cite{soft}, although its
practical relevance for $t \bar t$ production at
both colliders is still an issue of debate.
Soft gluon corrections and Coulomb gluon bound state corrections were combined
in \cite{Kiyo:2008bv} to provide an accurate description of $t \bar t$
threshold region.
Electroweak corrections to $\sigma_{t \tbar}$ were studied in \cite{ttbar:elweak}.
There is an ongoing effort to compute next-to-next-to-leading order QCD corrections to top quark pair production in hadron collisions  \cite{ttbar:nnlo}.

In all studies of top production described so far top quarks were treated
as stable particles and summation over their  spin degrees of freedom was
performed. This approach introduces two problems. First,
spin degrees of freedom of  top quarks influence kinematics of their
decay products and in this way lead to observable consequences.
Second, there are QCD radiative corrections related to the decay, rather
than production, stages of the time evolution of $t \bar t$ system.
 Given the fact that the
$t \bar t$ production cross-section at the LHC is very large, a degree
of realism in its description is clearly warranted. To achieve it,
 we require a
NLO  QCD prediction for $t \bar t$ production that is valid at the level
of observable particles, such as  leptons,
quarks and gluons originating {\it either} from production
of top quarks or in their decays. In principle,
this seems to necessitate  a next-to-leading  computation  for
 $2 \to 4$ processes such as
 $pp \to \ell^+ \nu \ell^- \bar \nu b \bar b$ or   $pp \to u \bar d \ell^-\bar \nu b \bar b$,
which is a formidable task at present.

Fortunately, the problem can be
simplified by studying  double resonance contributions, but
accounting for  spin degrees of freedom exactly through all
stages of the top quark  decay chain.
Indeed, when top quarks are treated as   truly unstable
particles, all QCD corrections to a relevant $2 \to 4$
process\footnote{We mean here the final state $W^+W^-b \bar b$.}
can be decomposed into factorizable and
non-factorizable \cite{my}.
Non-factorizable corrections imply a cross-talk
between production and decays of top quarks.  In the limit
$\Gamma_t /m_t \to 0$,  these corrections must vanish since quantum
interference  should not occur if events are separated by macroscopic
distances. The precise way in which such non-factorizable corrections vanish
was described in Refs.~\cite{my,Beenakker:1999ya,rc}.
In what follows we work in the
on-shell approximation for top quarks, ignoring
non-factorizable corrections.

Once non-factorizable corrections are neglected, a full description of
$t \bar t$ production and decay including all the spin correlations
is achieved by computing NLO QCD
corrections to both  production and decay of a {\it polarized} $t \bar t$
pair. In  an impressive series of papers
\cite{Bernreuther:2001bx,Bernreuther:2001rq,Bernreuther:2001jb}
Bernreuther, Brandenburg, Si and Uwer
computed the spin density matrix for the production of
a $t \bar t$ pair in hadron collisions through NLO QCD.
Corrections to decays of polarized top quarks
were obtained in
Refs.~\cite{Czarnecki:1990pe,Brandenburg:2002xr}.  Putting all these bits together,
Bernreuther et al. studied a number of kinematic distributions which
can be used at hadron colliders to probe top quark spin correlations
\cite{Bernreuther:2004jv,Bernreuther:2004wz}.

While the importance of accurate predictions of top quark spin correlations
was strongly emphasized in
Refs.~\cite{Bernreuther:2004jv,Bernreuther:2004wz},
to the best of our knowledge
there is no  publicly available  numerical
program that satisfies the following requirements:
\begin{itemize}
\item  it containts NLO QCD corrections to top quark
production and decay;
\item it   includes all spin correlations for the
processes
$pp \to \ell^+ \nu \ell^- \bar \nu b \bar b$  and
$pp \to u \bar d \ell^-\bar \nu b \bar b$;
\item it  allows arbitrary cuts on particles in the final state.
\end{itemize}
While the implementation of heavy flavor production in
MC@NLO \cite{mcnlo} and POWHEG \cite{powheg}
comes close to these requirements (see e.g. \cite{Frixione:2007zp}), those programs
do not fully include  all spin correlations through NLO QCD.
It seems to us that developing a numerical program with capabilities
listed above
is  useful since it will
contribute to more realistic description  of $t \bar t$ production
at hadron colliders.  Computing  NLO QCD corrections
to $t \bar t$ production and decay and implementing them into a flexible
numerical program    is the primary goal of this paper.

Another motivation for undertaking the study of $t \bar t$ production
is more theoretical -- we would like to explore
how unitarity-based methods for one-loop
computations work in a relatively simple  but fully realistic
 setting, when massive particles
are involved.  Initial studies
of  $t \bar t $ production in the context of
generalized $D$-dimensional  unitarity \cite{Giele:2008ve}
were performed
in \cite{Ellis:2008ir}.  Those results were extended by us with an eye on
applying generalized
$D$-dimensional unitarity to $t \bar t$ production in association
with jets\footnote{We point out that
 NLO QCD corrections to $t \bar t +{\rm jet}$
production at the LHC and the Tevatron were reported in
\cite{Dittmaier:2007wz,
Dittmaier:2008uj}.
}.  The simplest application -- the case of polarized
$t \bar t$ production -- is described  in this paper.

The rest of the paper is organized as follows. In Section~\ref{sect2}
some theoretical aspects of the computation are discussed. In
Section~\ref{sect3} we present a number of observables relevant for
$t \bar t $ studies at the Tevatron and the LHC. We conclude in
Section~\ref{sect4}.
More details on the calculation of virtual and real corrections are described in
Appendices \ref{appA} and \ref{appB}, respectively.

\section{Theoretical framework}
\label{sect2}

In this Section the theoretical framework relevant for the  computation is
briefly described. As already pointed out in the Introduction, we study the
production of top quarks and allow for their semileptonic
decays. We  include all the
spin correlations but  keep top quarks on their  mass shells.
We do not consider hadronic decays of top quarks in this paper.

We first discuss an efficient way to incorporate top quark decays into
the computation\footnote{We are grateful to Keith Ellis for emphasizing this point to us.}.
The tree amplitude for the process $ij
\rarr \tbar t
\rarr (\bar{b} \ell^- \bar \nu)(b \ell^+ \nu)$ can be written as
\beq
\label{eqn:decayamp1}
   A^{\tree} = \Bigg( \tilde A(t\rarr b \ell^+ \nu ) \; \frac{\ri(\Fslash{p}_t+m_t)}{p_t^2-m_t^2+\ri m_t \Gamma_t}  \Bigg)
       \; \tilde A(ij \rarr \tbar t) \;
       \Bigg( \frac{\ri(-\Fslash{p}_{\tbar}+m_t)}{p_{\tbar}^2-m_t^2+\ri m_t \Gamma_t} \; \tilde A(\tbar \rarr \bar{b} \ell^- \bar \nu) \Bigg),
\eeq
where $\tilde A(ij \!\rarr\! \tbar t)$ and
$\tilde A( \tparbar \!\rarr\!\bparbar \ell^\pm \nuparbar)$
are sub-amplitudes for production and decay processes,
top quark propagators are factored out and summation
over spinor indices is implicit.
Note that in Eq.(\ref{eqn:decayamp1})
the top quarks are off the mass-shell.
To compute the cross-section, we need to square the amplitude $A^{\rm tree}$
and integrate it over the phase-space for final state particles.
To  simplify this procedure,  we approximate the squared amplitude
by taking the limit $\Gamma_t/m_t \rarr 0$. In this approximation,
propagators that appear in \refeq{eqn:decayamp1} yield
\beq
\label{eqn:nwaapprox}
   \frac{1}{(p_t^2-m_t^2)^2 + m_t^2\Gamma_t^2}
{\Bigg |}_{\Gamma_t/m_t \to 0} =
\quad \quad \frac{2\pi}{2m_t\Gamma_t} \; \delta (p_t^2-m_t^2).
\eeq
After factorizing the phase-space integral into the phase-space for a
$t \bar t$  pair and the decay phase-spaces for $t$ and $ \bar t$,
the delta-function in Eq.(\ref{eqn:nwaapprox})
forces   top quarks on their mass-shells.
This separates the production stage from the decay stage and allows
us to implement  top quark  decays by choosing the on-shell spinors
\beq
\label{eqn:topspi}
   \bar{U}(p_t) &=& \tilde A(t\rarr b \ell^+ \nu) \; \frac{\ri(\Fslash{p}_t+m_t)}{\sqrt{2 m_t \Gamma_t}},
   \\
\label{eqn:topspi2}
   V(p_{\tbar}) &=& \frac{\ri(-\Fslash{p}_{\tbar}+m_t)}{\sqrt{2 m_t \Gamma_t}} \; \tilde A(\tbar\rarr \bar{b} \ell^- \bar \nu),
\eeq
to describe polarization states of top quarks.
\refeq{eqn:decayamp1} can now be rewritten as
\beq
\label{eqn:decayamp2}
  A^{\tree} = \bar{U}(p_t) \; \tilde A(ij\rarr \tbar t) \; V(p_{\tbar}) + \mathcal{O}\left(\frac{\Gamma_t}{m_t}\right).
\eeq
The usefulness of \refeqs{eqn:topspi}{eqn:decayamp2} is twofold.
First, since  $\bar{U}$ and $V$ are on-shell Dirac spinors, the amplitude
$A^{\rm tree}$ can be computed in a conventional way;
this  means that the inclusion
of $t$ and $\bar t$
decays does not increase the complexity of scattering
amplitudes that
need to be calculated.
Second, \refeq{eqn:decayamp2} actually helps in
{\it reducing} computational burden.
Indeed, the top quark spinors depend on polarization
states and momenta of leptons, neutrinos and $b$-quarks which
we treat as massless.
For top quark decays,
helicity states of all massless particles are  fixed,
due to the $V-A$ structure of  flavor-changing interaction
vertices in the Standard Model.
This implies that the spinors $\bar{U}$ and $V$ have {\it uniquely defined polarizations} and no helicity sums related to $t$ and ${\bar t}$ are involved in the cross-section computation.
In the evaluation of the decay amplitudes we keep  the $W$-boson
on-shell and treat the $b$-quark as massless particle.

Having discussed how to deal with
$t$ and $\bar t$ decays efficiently,
we focus on other computational details.
The on-shell approximation for top quarks
splits all the QCD effects into  corrections to the production and corrections
to the decay. These corrections do not interfere since production and
decay stages are separated by large space-time intervals. As the result,
$t \bar t$ production with NLO QCD effects and $t(\bar t)$ decay
with NLO QCD effects can be treated as independent  processes.

We begin with corrections to $t \bar t$  production process.
In this case, as we just discussed, all
one has to do to incorporate  top quark decays, is to use
in Eq.(\ref{eqn:decayamp2})
the on-shell spinors for $t$ and $\bar t$, shown in
 \refeqs{eqn:topspi}{eqn:topspi2}.
This prescription is valid both at leading  and
at next-to-leading order. We organize the  calculation in terms of
gauge-invariant color ordered sub-amplitudes.
Each such amplitude multiplies a specific color factor and has a fixed ordering of external particles.
We use Berends-Giele \cite{BG} recurrence relations to evaluate
those sub-amplitudes for each contributing helicity configuration at
leading order.
There are two partonic processes that need to be considered for LO, and
virtual component of NLO computations --
$gg \to t \bar t$ and $q \bar q \to t \bar t$.
Their color decomposition is given in Appendix~\ref{appA}.

For the computation of virtual corrections to $t \bar t$  production,
we  employ the method of
 generalized $D$-dimensional unitarity \cite{Giele:2008ve}.
The basic idea of this method is to consider unitarity cuts of
scattering amplitudes  in  $D$-dimensional space-times where $D>4$
is {\it integer}.
It turns out that dimensionally-regularized
one-loop amplitudes can be fully reconstructed from
these unitarity cuts provided that
{\it tree-level}
 on-shell scattering
amplitudes for  complex external momenta, evaluated
in  six- and eight-dimensional space-times, are available.
For a detailed description of the method see Ref.~\cite{Giele:2008ve}.

We turn to the discussion of  real emission corrections to
$t \tbar$ production.
Real emission processes develop singularities when a massless
final state particle is unresolved, i.e.
it is soft and/or collinear to another particle.
When integrated over the unresolved phase-space, those
 singularities produce divergencies that   cancel against
divergencies   in  virtual corrections
and unrenormalized parton distribution functions,
yielding finite cross-sections.
In order to treat real emission singularities in a numerically
 stable way we employ an extension of
Catani-Seymour dipole subtraction scheme \cite{Catani:1996vz}
to massive particles \cite{Catani:2002hc}. Recall that the
 basic idea of the subtraction
procedure  is to construct an   approximation to matrix elements squared
of real emission processes that, on one hand, has the same singular
limits as real emission matrix elements  and, on the other hand,
can be analytically
integrated over the phase-space of unresolved particles. Because
top quarks are heavy, they never appear as unresolved
particles and their spin degrees of freedom are not essential for the
construction of the subtraction terms.
Hence, we can also use the ${\bar U},V$ spinors to compute (subtracted)
real emission
corrections and in this way account for decays of top quarks exactly.
The color decomposition of amplitudes needed for the real corrections to the $t \tbar$ production process is given in Appendix~\ref{appA}.
Furthermore, a list of all dipole subtraction terms is given in Appendix~\ref{appB}.

We now turn to the discussion of NLO QCD corrections to the top quark decay.
Corrections to decay rate and lepton kinematic distributions in  decays of polarized top quarks  are well-known \cite{topwidth,Czarnecki:1990pe}.
However, since we want to keep our computation completely differential, to allow for arbitrary  cuts on the final state particles, we need to go beyond the computation of Ref.\cite{Czarnecki:1990pe}.
A detailed discussion of how such a computation should be set up was given recently for the case of single top production in Ref.\cite{Campbell:2004ch} and
we closely follow that reference  in our implementation of virtual and real QCD corrections
to top quark decays.

We also point out that  we use $\Gamma_t^{\rm LO}$ and $\Gamma_t^{\rm NLO}$  to construct top quark
spinors in Eqs.~(\ref{eqn:topspi})-(\ref{eqn:topspi2}) in
LO  and NLO  computations, respectively.
Indeed if no restrictions on final state particles are applied, the cross-section computed in the on-shell approximation is given by the product of the $t\bar t$ production cross-section multiplied by $t$ and $\bar t$ decay branching fractions.
Since this statement holds true to all orders in perturbation theory, the width in  \refeqs{eqn:topspi}{eqn:topspi2}  should also be computed in series of $\alpha_s$, for consistency.
Hence, the complete NLO cross-section can schematically be written as
\beq
\label{eqn:nloxsec}
 \rd\sigma^\mathrm{NLO} &=& \rd\sigma^0 \; \frac{\rd\Gamma^0_{\tbar} \; \rd \Gamma^0_t}{(\Gamma^0)^2}
    \;+\; \rd\sigma^1 \; \frac{\rd\Gamma^0_{\tbar} \; \rd \Gamma^0_t}{(\Gamma^0)^2}
    \nl
    && \;+\; \rd\sigma^0 \; \left( \frac{\rd\Gamma^1_{\tbar} \; \rd \Gamma^0_t}{(\Gamma^0)^2} + \frac{\rd\Gamma^0_{\tbar} \; \rd \Gamma^1_t}{(\Gamma^0)^2} \right)
    \;- \; \rd\sigma^0 \;  \frac{2 \Gamma^1_t}{\Gamma^0_t} \; \frac{\rd\Gamma^0_{\tbar} \; \rd \Gamma^0_t}{(\Gamma^0)^2}
\eeq
where $\sigma^0,\Gamma^0$ and  $\sigma^1,\Gamma^1$ denote leading
and next-to-leading order contributions to the production and
decay processes, respectively.
The last term in \refeq{eqn:nloxsec}
arises from the correction to the top quark
width in Eqs.(\ref{eqn:topspi})-(\ref{eqn:topspi2}).\\

To conclude this Section, we mention checks that
we applied to our computation to ensure its correctness.
Within the unitarity
method, one computes  scattering amplitudes for
particular polarization states of external particles.
As the result, gauge invariance of all one-loop amplitudes involving external gluons
can be easily checked by substituting polarization vector of a gluon
by its momentum. One can also check the threshold
$s \to 4 m_t^2$  behavior of one-loop amplitudes, relevant for $t \bar t$
production.  Indeed, close to
$t \bar t$ threshold, scattering amplitudes are dominated by
the Coulomb singularity
and we checked that our implementation of the virtual corrections
reproduces the Coulomb singularity correctly.
The cancellation of infra-red and collinear divergencies
 between virtual and
real corrections was checked numerically.
Furthermore, the implementation of both virtual and real corrections was checked
by comparing results for physical observables produced by our code
against other programs that compute the $t \bar t$ production cross-section without top decays.
We have extensively used MadGraph \cite{Stelzer:1994ta} and MCFM \cite{Ellis:2006ar} for these comparisons
to check LO and NLO results, respectively.
In order to ensure the correctness of our implementation of top quark decays we
checked that the NLO QCD corrections reproduce known results for the top width if no
restrictions on final state particles are applied.  Finally, we have checked
that our code reproduces NLO QCD corrections to
some  observables sensitive to top quark
spin correlations, discussed in Ref.\cite{Bernreuther:2004wz}.

\section{Results}
\label{sect3}

We have implemented NLO QCD corrections to the production and semileptonic
decays of a top quark pair in a FORTRAN program. The goal of this
Section is to illustrate some of its potential applications. Before discussing
those examples, we describe the input parameters that are used in all numerical
results presented in this paper.

We  consider production of $t \bar t$ pairs both at the Tevatron
and the LHC. The Tevatron center of mass energy is $\sqrt{s} = 1.96~{\rm TeV}$
and the LHC center of mass energy is $\sqrt{s} = 10~{\rm TeV}$.
We use the pole mass of the top quark, $m_t = 172~{\rm GeV}$. The strong
coupling constant $\alpha_S$ is renormalized in the ${\overline {\rm MS}}$
scheme except for top quark loop contributions to gluon
self-energy diagrams that are subtracted at zero gluon momentum.
We employ  CTEQ6L1 parton distribution functions for
leading order and CTEQ6.1M for next-to-leading order computations
\cite{Pumplin:2002vw,Nadolsky:2008zw}; those PDF sets
correspond to $\alpha_s(m_Z) = 0.130$ and $\alpha_s(m_Z) = 0.118$
respectively.   The mass of the $W$ is taken to be $m_W
= 80.419~{\rm GeV}$. $W$ couplings to fermions are obtained from
the Fermi constant $G_\mathrm{F} = 1.16639 \cdot 10^{-5} \,{\rm GeV}^{-2}$.
The leading order width of the top quark is $\Gamma_t^{\mathrm{LO}} = 1.47~{\rm GeV}$
and $\Gamma_t^{\mathrm{NLO}} = 1.31~{\rm GeV}$ at next-to-leading order\footnote{This result for  $\Gamma_t^{\rm NLO}$ is obtained by choosing
the renormalization scale for the strong coupling constant to be the top quark mass.}.
We use $|V_{tb}| = 1$ and employ on-shell
approximation for the $W$ boson, produced in  $t \to W b$ decays.

We set the width of the $W$ boson
to $2.14~{\rm GeV}$. We consider this to be an experimentally measured
width of the $W$ boson, not the result of computation in leading order of
QCD perturbation theory.  This implies that our leading order prediction
for $pp \to t \bar t \to \ell^+ \nu \ell^- \bar \nu b \bar b$ includes the NLO
QCD  branching ratio for $W \to e \nu $. While simple, this effect is
definitely not negligible numerically --  since
 ${\cal O}(\alpha_s)$ corrections increase the
 $W$ width by  approximately $2.5-3$
percent, the use of LO or NLO branching fractions for $W$ bosons changes
the cross-section for $pp \to t \bar t \to \ell^+ \nu \ell^- \bar \nu b \bar b$
by five to six  percent.

\begin{figure}[!t]
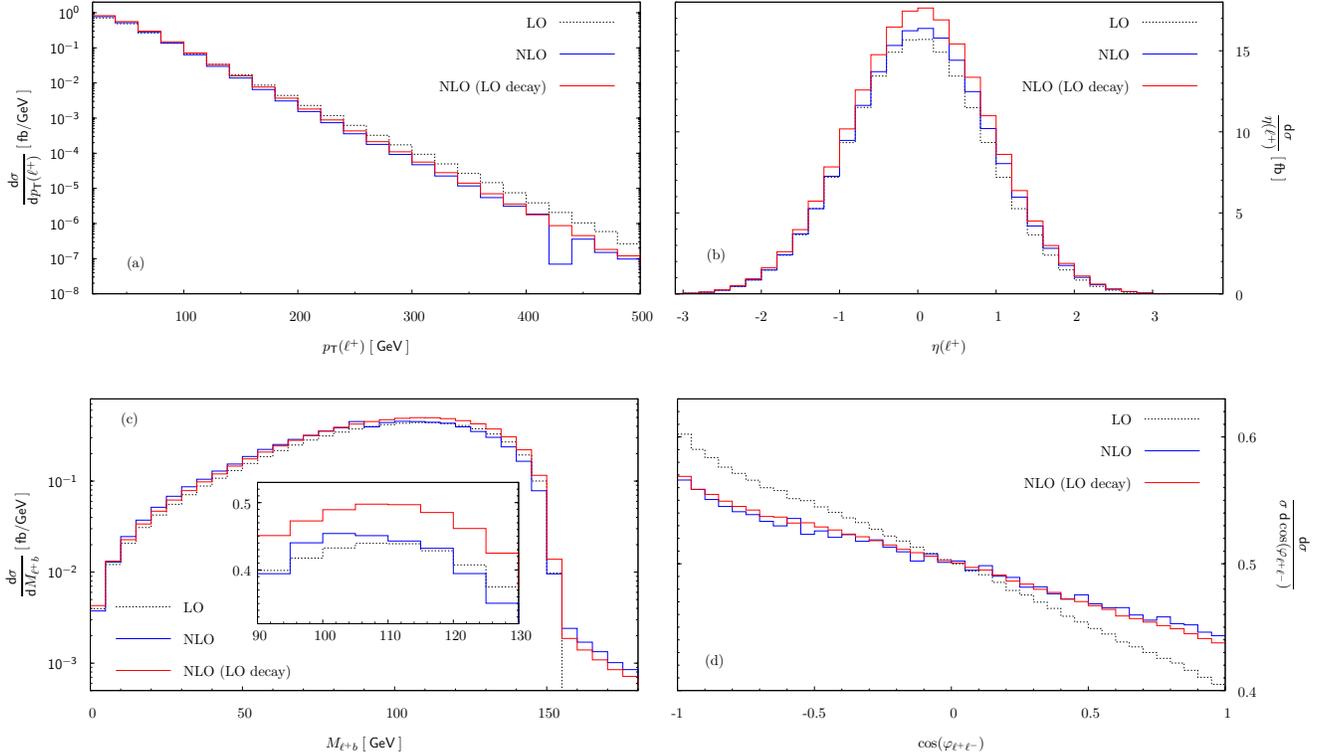

 \begin{center}
  \scalebox{0.5}{\input{TEV_Fig01.tex}}
  \scalebox{0.5}{\input{TEV_Fig02.tex}} \\[8mm]
  \scalebox{0.5}{\input{TEV_Fig03.tex}} \hspace{12mm}
  \scalebox{0.5}{\input{TEV_Fig04.tex}}
 \end{center}
\caption{
Various kinematic distributions for leptonic final
states in $t \bar t$ production at the Tevatron. We show transverse
momentum (a) and rapidity (b) distributions of the positively charged lepton as well
as the distribution in the invariant mass of a lepton and a $b$-jet (c).
The distribution in the (particularly defined, see text)
opening angle of the leptons is shown in (d).
Each distribution in panel (d) is normalized to
the corresponding total cross-section.
All cuts described at the beginning of Section III are applied.
 }
\label{fig1}
\end{figure}

To define jets,  we use the $k_\perp$-clustering
algorithm \cite{Catani:1992zp} with $R = 0.4$. We require that two $b$-jets are present in the
event.  We define jet flavor through its  ``bottomness''  quantum number so
that, for example, a jet that contains  $b$ and $\bar b$
is not a $b$-jet. We point out that
this definition  is infra-red safe through NLO QCD approximation
for $t \bar t$ production whereas this is not true  in general, for massless
quarks \cite{Banfi:2006hf}.
For $b$-jets we require a minimal transverse momentum of $20~{\rm GeV}$.
Charged leptons must be produced with the  transverse momentum larger than
$20~{\rm GeV}$ and the missing energy in the event should
exceed $40~{\rm GeV}$.
We also set renormalization and factorization scales to the value of the
top quark  mass.
Since scale dependence of various observables for $t \bar t$ production
was studied in detail in the existing literature,
we do not present such studies in this paper.

\begin{figure}[!t]
 \begin{center}
  \scalebox{0.52}{\input{LHC_Fig01.tex}}
  \scalebox{0.52}{\input{LHC_Fig02.tex}} \\[8mm]
  \scalebox{0.52}{\input{LHC_Fig03.tex}} \hspace{12mm}
  \scalebox{0.52}{\input{LHC_Fig04.tex}}
 \end{center}
\caption{
Various kinematic distributions for leptonic final
states in $t \bar t$ production at the LHC. We show transverse
momentum (a) and rapidity (b) distributions of the positively charged lepton as well
as the distribution in the invariant mass of a lepton and a $b$-jet (c).
The distribution in the (specially defined, see text)
opening angle of the leptons is shown in (d).
Each distribution in panel (d) is normalized to
the corresponding total cross-section.
All cuts described at the beginning of Section III are applied.
 }
 \label{fig2}
\end{figure}

To set the scale for the magnitude of next-to-leading QCD effects in $t \bar t$
production for our choices of input parameters,
we quote results for cross-sections at leading and
next-to-leading order for the Tevatron $(p \bar p \to t \bar t \to
bl^+\nu  \bar b l^- \bar \nu)$
\be
\sigma_{\rm LO}  = 34.63 \, \mathrm{fb},\quad \sigma_{\rm NLO} = 36.47 \, \mathrm{fb},\;\;\;
K_{\rm TEV} = \frac{\sigma_{\rm NLO}}{\sigma_{\rm LO}}=1.05,
\ee
and the LHC $(p p \to t \bar t \to
bl^+\nu \bar b l^- \bar \nu)$
\be
\sigma_{\rm LO}  = 1484 \, \mathrm{fb}, \quad \sigma_{\rm NLO} = 2097 \, \mathrm{fb},\;\;
K_{\rm LHC} = \frac{\sigma_{\rm NLO}}{\sigma_{\rm LO}}=1.41.
\ee
To obtain those numbers, we set the  renormalization and factorization scales
to $m_t$ and apply all the cuts listed in the beginning of this Section.

We are now in position to illustrate capabilities
of our numerical program by presenting a number of  $t \bar t$-related
 kinematic distributions,  computed
through NLO in perturbative QCD.
In Fig.\ref{fig1} we present results for the Tevatron.
The transverse momentum and rapidity distributions of leptons in top decays are shown in Figs.~\ref{fig1}(a) and \ref{fig1}(b), respectively.
The distribution in the invariant mass of the charged lepton and the $b$-jet is given in Fig.\ref{fig1}(c).
Fig.\ref{fig1}(d) shows the distribution in $\cos\varphi_{\ell^+\ell^-}$, where $\varphi_{\ell^+\ell^-}$ is the angle between the directions
of flight of $\ell^+$ and $\ell^-$,
defined in the rest frames of $t$ and $\bar t$ respectively.
In all cases we compare predictions at leading and next-to-leading order, with and without corrections to the decay.
Corresponding results for the LHC are shown in Fig.\ref{fig2}.

\begin{figure}[!t]
 \begin{center}
\scalebox{0.5}{\input{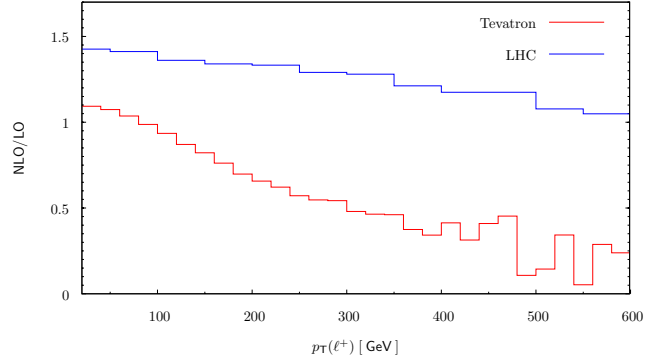}}
 \end{center}
\caption{
The ratio of NLO  to LO predictions for $\ell^+$  transverse momentum
distributions, for the Tevatron and the LHC. The
input parameters are described at the beginning of Section III.
 }
\label{fig4}
\end{figure}

We first consider transverse
momentum and rapidity distributions of the charged lepton, shown in the upper
panels of  Figs.\ref{fig1},\ref{fig2}. Both of these distributions are
standard but effects of QCD corrections to top  decays and top
spin correlations are never included in their computation.
For both of these observables NLO QCD effects are important.
For example, as we show in Fig.\ref{fig4}, the shape
of the transverse momentum distribution changes significantly at the Tevatron
while at the LHC similar change is smaller, but non-negligible.
For the charged lepton
rapidity distribution, NLO QCD corrections to decays are  important,
especially at central rapidities at the Tevatron.

Another distribution that we show in Figs.\ref{fig1},\ref{fig2}
is the distribution in the invariant mass of a positively charged
lepton $\ell^+$ and a $b$-jet, $M_{\ell^+b}$.
This observable is  interesting for a number of
reasons. First, a (kinematically)  similar observable -- an invariant mass
of $\ell^+$ and $J/\psi$ originating from $B$ meson decay --
was discussed in connection with the measurement of the top quark
mass in Ref.\cite{Kharchilava:1999yj}.  Similar to $J/\psi\;\ell^+$ invariant mass,
the invariant mass of the $b$-jet and $\ell^+$
has a clear kinematic boundary at leading order,
see Figs.\ref{fig1},\ref{fig2}.
This boundary ${\rm max} (M_{\ell^+b}^2) = m_t^2 - m_W^2$
is the consequence of the on-shell conditions for the
top quark and the $W$ boson. Those conditions lead to a
lower bound on the allowed neutrino energy in the top quark rest frame $E_\nu$
and to the upper bound on $M_{\ell^+b}^2 = m_t^2 - 2m_t E_{\nu}$.
The existence of the kinematic  boundary
makes this observable potentially interesting for the top quark
mass determination.
In addition,
$M_{\ell^+b}$ belongs to a general class of observables that may be used
to study spins of Beyond the Standard Model particles at the LHC. The effect
of NLO QCD corrections on the $M_{\ell^+b}$ distribution is interesting.
For example, at the Tevatron there is a significant cancellation between
corrections to $t \bar t$ production and corrections to $t$ and $\bar t$
decays for this observable. This effect also exists at the LHC although
it is less pronounced.

It is interesting that once the NLO QCD
corrections are included, events start to appear beyond the leading order
kinematic boundary and both, QCD corrections to the production and QCD
corrections to the decay contribute. The origin of these events
is peculiar since they appear as the consequence  of the fact that
the momentum of the $b$-jet rather than the momentum of the
$b$-quark is employed in the calculation of $M_{\ell^+b}$. Then, for example,
a gluon emitted in the decay of $\bar t$ may combine with the $b$-quark
from top decay to form a $b$-jet. The energy of such a jet is not restricted
by on-shell conditions applied to $t$ and $W$. As the result, the invariant
mass of the $b$-jet and $\ell^+$ may exceed the LO kinematic boundary. A similar
mechanism is in effect when gluon emission occurs as part of the top production
process. Note, however, that if a gluon is emitted in the decay of a {\it top}
quark  $t \to b \ell^+ \nu +g$, then the invariant mass of a $b$-jet and $\ell^+$
can not exceed the leading-order kinematic boundary.

Finally, we  discuss lepton angular correlations; such observables
are particularly sensitive to correct implementation of spin correlations
in $t \bar t$ production and decay. Indeed, angular correlations \
tell us  whether leptons prefer
to be produced with parallel or anti-parallel momenta and, as it turns
out, the answer to this question differs for the Tevatron and the LHC
\cite{bernr}.
To understand the difference
recall the two  mechanisms that dominate
the  $t \bar t$ pair production
at those colliders. At the Tevatron, top pairs
are mostly produced in an annihilation of a
$q \bar q$ pair -- $q \bar q \to g^* \to t \bar t$
-- which forces a $t \bar t$ pair to have
angular momentum $J=1$. Since, in addition, most of the time
 the annihilation occurs close
to $t \bar t$ threshold, it is most probable that the
$t \bar t$ pair is  produced in an $S$-wave with  spins of
$t$ and $ \bar t$ parallel, to create a $J=1$ state.
Since $e^+$ likes to follow the direction of
$t$ spin while $e^-$ prefers the direction opposite to $\bar t$
spin,  flight directions of $e^+$ and $e^-$
are anti-correlated.

At the LHC the situation is different since
$gg \to t \bar t$ annihilation becomes the dominant
production channel.  Close to $t \bar t$ threshold
 the largest contribution  comes
from an $S$-wave $J=0$ color-octet annihilation \cite{Kiyo:2008bv}.
This suggests that
$e^+$ and $e^-$ prefer to be produced with parallel
momenta at the LHC, at least to an extent
that  threshold production
is important there.

In  Figs.~\ref{fig1}(d),\ref{fig2}(d) we show  the
opening angle distribution between
two charged leptons at the LHC and the Tevatron. Note that
the flight directions of $\ell^+$ and $\ell^-$ are defined, respectively,  in
$t$ and $\bar t$ {\it rest} frames, when computing this observable.
Azimuthal  correlations are substantial at LO and remain fairly
pronounced even after NLO QCD corrections are included. It is interesting
that NLO QCD effects are more important at the Tevatron where the shape
of the distribution changes. At the LHC
the NLO QCD effects  do not change the shape of the distribution at all, so
leading order predictions do a good job in that case.  QCD corrections
to top decays do not play an important role for this observable for both,
the Tevatron and the LHC. We note that a similar distribution was computed
through NLO QCD in Refs.\cite{Bernreuther:2004jv,Bernreuther:2004wz}, but
no jet cuts were applied.  However, by
removing all the cuts in our computation
and by adjusting the input parameters, we reproduce results for this
distribution reported
in Refs.\cite{Bernreuther:2004jv,Bernreuther:2004wz}.

\begin{figure}[!t]
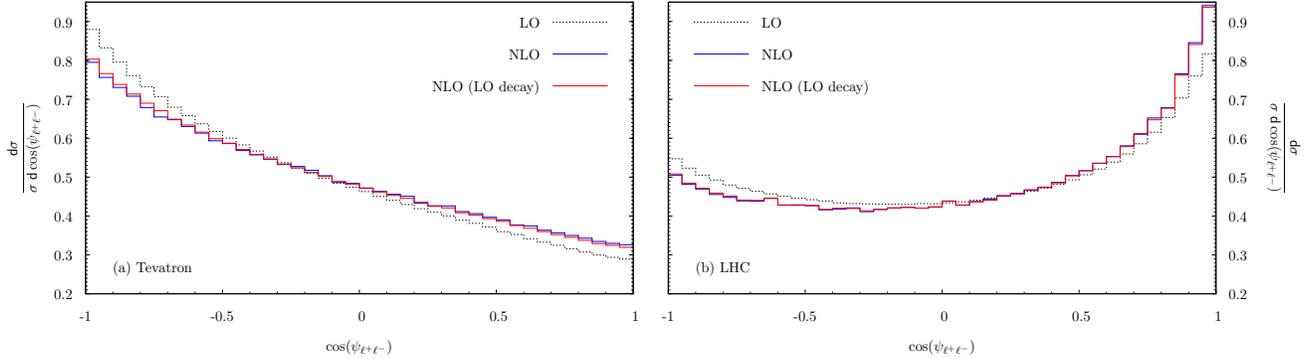

 \begin{center}
\scalebox{0.5}{\input{TEV_Fig05.tex}}
\scalebox{0.5}{\input{LHC_Fig05.tex}}
 \end{center}
\caption{
The distribution in the
opening angle of the two leptons in the laboratory frame,
normalized to
the corresponding total cross-section.
All cuts described at the beginning of Section III are applied.
 }
\label{fig25}
\end{figure}

One of the drawbacks of using  $\cos \varphi_{\ell^+\ell^-}$ to
study spin correlations
is the requirement
that $t$ and $\bar t$ rest frames are
reconstructed  and, for
semileptonic top decays,  it is
not possible to do this unambiguously. In this respect, we
would like to point out that $t \bar t$ spin correlations can also be
studied using a much simpler observable -- the opening angle $\psi_{\ell^+\ell^-}$
of the two leptons
in the {\it laboratory frame}. The corresponding distributions
computed at leading and next-to-leading order  are shown in
Fig.\ref{fig25} for both the Tevatron and the LHC. The results are
similar to distributions in $\varphi_{\ell^+\ell^-}$ shown in
Figs.~\ref{fig1}(d),\ref{fig2}(d) but the effect of the radiative
corrections on the shape of $\psi_{\ell^+\ell^-}$ distribution at the LHC
is stronger.

To conclude this Section, we point out that
results reported here are not supposed
to provide  exhaustive phenomenological studies of $t \bar t$ production;
rather, they should illustrate capabilities of our numerical implementation
of NLO QCD corrections to $t \bar t$  production and decays.
We have chosen to present a number of distributions which we find
interesting  but, as with any NLO QCD computation, essentially any infra-red
safe observable can be calculated and analyzed. In this regard,
a numerical program that allows to impose
arbitrary cuts and  employ arbitrary
jet algorithms, includes all the spin correlations in $t \bar t$ production
and in $t$ and $\bar t$  decays  and incorporates  NLO QCD corrections
to (semileptonic) decays of top quarks, is very attractive since it can be
used for realistic description of  $t \bar t$ pair production
in hadron collisions.

\section{Conclusions}
\label{sect4}

In this paper the computation of
next-to-leading order QCD corrections to the production
and decay of top quark pairs at the Tevatron and the LHC
is presented. Our goal was to  develop a numerical
program which employes  the on-shell approximation for top quarks,
yet accounts for  all the spin correlations
in their production and semileptonic decays through NLO QCD.

To this end, the  implementation of next-to-leading order
corrections to the production and  decay of polarized top quarks is required.
We performed such an implementation at a fully differential level.
This allows us to compute an arbitrary observable defined in terms  of
lepton momenta, missing energy and jet energies and momenta,
originating either in top production or decay. We have illustrated the
capabilities of the program by presenting a  variety of differential
distributions that are sensitive to top spin correlations and NLO QCD
corrections to top quark decays.

An interesting aspect of the  computation reported in this paper is that
the method of
generalized $D$-dimensional  unitarity \cite{Giele:2008ve} is employed
to compute virtual corrections to $t \bar t$ production cross-section;
this is the first application
of generalized $D$-dimensional
unitarity to a fully realistic NLO computation that
involves massive external and internal particles.
We find that, in general,  the method works well  so its application
to other processes with massive particles is definitely warranted.

\vspace*{0.2cm}
{\bf Acknowledgments}
We are grateful to Keith Ellis for useful conversations.
This research  is supported by the startup package provided by
Johns Hopkins University.

\begin{appendix}

\section{}
\label{appA}

In this Appendix we summarize various results for the color decomposition
of tree- and one-loop amplitudes that we used in this paper.

We begin with the leading order process. At leading order  two partonic
processes contribute to $t\bar t$ production: $gg\rarr t\tbar$ and $\qbar q \rarr t\tbar$.
Their color decomposition is given by
\beq
\label{eqn:coltreegg}
   A^\tree(gg\rarr t\tbar) &=& \gs^2
   \sum_{\sigma \in S_2} (T^{a_{\sigma_3}} T^{a_{\sigma_4}})_{i_2}^{\bar{i}_1} \; \calA^\tree (1_{\tbar}, 2_t,(\sigma_3)_g,(\sigma_4)_g),
\\
\label{eqn:coltreeqqb}
   A^\tree(\qbar q \rarr t\tbar) &=& \gs^2 \Big[
   \delta^{\bar{i}_1}_{i_4} \delta^{\bar{i}_3}_{i_2}
  -\frac{1}{\Nc} \delta^{\bar{i}_1}_{i_2} \delta^{\bar{i}_3}_{i_4}
   \Big]
  \calA^\tree(1_{\tbar}, 2_t,3_{\qbar},4_q)
\eeq
where $\calA^\tree(1_{\tbar},2_t, i,j)$ are color-ordered tree amplitudes and $\gs$ is the strong coupling constant.
The generators $T^a$ of the SU($\Nc=3$) color group are normalized to $\mathrm{Tr}(T^aT^b)=\delta^{ab}$ and satisfy the commutation relation
$[T^a,T^b]=-F^c_{ab}T^c$.

Considering real emission contribution  to top quark pair production
at next-to-leading order,
we find that four partonic processes contribute
$gg \rarr \tbar t g$, $\qbar q \rarr \tbar t g$, $q g \rarr \tbar t q$, $\qbar g \rarr \tbar t \qbar$.
The last three processes are related by crossing symmetry.
The color decomposition for the two master processes is given by
\beq
   A^\tree(gg\rarr t\tbar g) =& \gs^3 &
   \sum_{\sigma \in S_3} (T^{a_{\sigma_3}} T^{a_{\sigma_4}} T^{a_{\sigma_5}})_{i_2}^{\bar{i}_1} \; \calA^\tree (1_{\tbar}, 2_t,(\sigma_3)_g,(\sigma_4)_g,(\sigma_5)_g),
   \\
   A^\tree(\qbar q \rarr t\tbar g) =& \gs^3 \Big[&
   (T^{a_5})^{\bar{i}_1}_{i_4} \; \delta^{\bar{i}_3}_{i_2} \; \calA^\tree(1_{\tbar},2_t,3_{\qbar},4_q,5_g)
   \nl
   && + (T^{a_5})^{\bar{i}_3}_{i_2} \; \delta^{\bar{i}_1}_{i_4} \; \calA^\tree(1_{\tbar},2_t,5_g,3_{\qbar},4_q)
   \nl
   && +\frac{1}{\Nc} (T^{a_5})^{\bar{i}_1}_{i_2} \; \delta^{\bar{i}_3}_{i_4} \; \calA^\tree(1_{\tbar},5_g,2_t,3_{\qbar},4_q)
  \nl
   && +\frac{1}{\Nc} (T^{a_5})^{\bar{i}_3}_{i_4} \; \delta^{\bar{i}_1}_{i_2} \; \calA^\tree(1_{\tbar},2_t,3_{\qbar},5_g,4_q)
  \Big].
\eeq

In the case of virtual amplitudes, a decomposition into color-ordered
amplitudes is insufficient; instead, one has to consider the so-called
primitive amplitudes \cite{Bern:1994fz}. We need one-loop amplitudes
for $gg \to t \bar t$ and $q \bar q \to t \bar t$. For $gg \to t \bar t$
the color decomposition reads
\beq
\label{eqn:col1Lgg}
  A^\virt(gg\rarr t\tbar) = \gs^4
  \sum_{\sigma \in S_{2}} &\Big[ &
  ( T^{x_2} T^{x_1} )_{i_2}^{\bar{i}_1} (F^{a_{\sigma_{4}}} F^{a_{\sigma_3}})_{x_1 x_2}
  \; \calA^{\rL,[1]}(1_{\tbar},2_t,(\sigma_3)_g,(\sigma_4)_g)
  \nl
  && + ( T^{x_2} T^{\sigma_3} T^{x_1} )_{i_2}^{\bar{i}_1} (F^{a_{\sigma_{4}}})_{x_1 x_2}
  \; \calA^{\rL,[1]}(1_{\tbar},(\sigma_3)_g,2_t,(\sigma_4)_g)
  \nl
  && + ( T^{x_1} T^{\sigma_4} T^{\sigma_3} T^{x_1} )_{i_2}^{\bar{i}_1}
  \; \calA^{\rL,[1]}(1_{\tbar},(\sigma_3)_g,(\sigma_4)_g),2_t)
  \nl
  &&+\sum_{f=1}^{\Nf} (T^{a_{\sigma_3}} T^{a_{\sigma_4}})_{i_2}^{\bar{i}_1} \; \calA^{\rL,[1/2]}_f (1_{\tbar}, 2_t,(\sigma_3)_g,(\sigma_4)_g)
  \Big]
\eeq
where $\calA^{\rL,[s]}$ are left-primitive amplitudes, defined in \cite{Bern:1994fz}.
$\Nf$ is the number of quark flavors; top quarks in closed fermion loops are treated as massive particles.
\\
\begin{figure}[!t]
\begin{center}
\epsfig{file=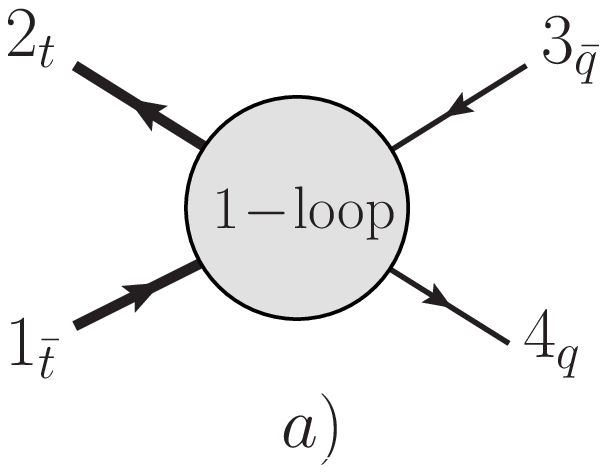, angle=0,width=0.2\textwidth}
\hspace{1cm}
\epsfig{file=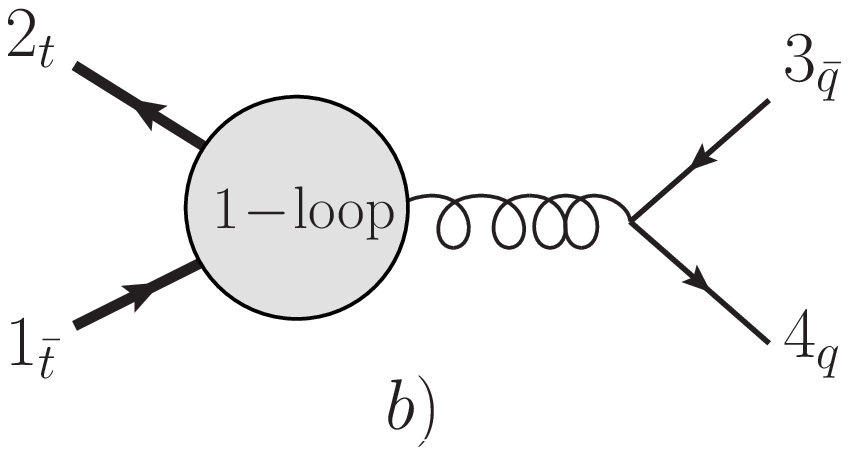, angle=0,width=0.27\textwidth}
\\[0.3cm]
\epsfig{file=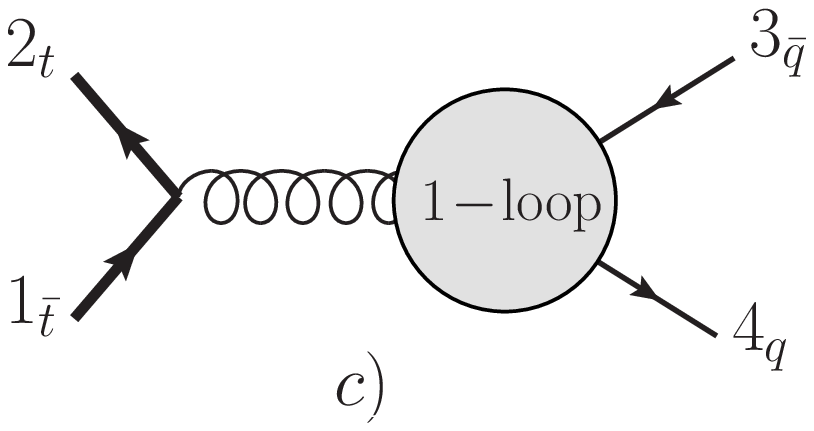, angle=0,width=0.27\textwidth}
\hspace{1cm}
\epsfig{file=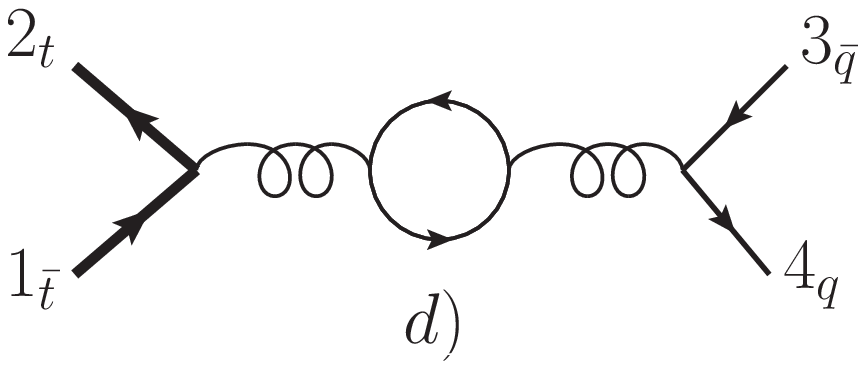, angle=0,width=0.29\textwidth}
\caption{Definition of primitive amplitudes with four quarks.
      Fig.~a) defines $\calA^{a}$ which includes all topologies where both
external fermion lines enter the loop;
    Figs.~b) and c) define $\calA^{b}$ and $\calA^{c}$ where either the top quark or the light quark line enter the loop, respectively;
      Fig.~d) $\calA^{d}$ includes topologies with a closed fermion loop.}
\end{center}
\label{fig35}
\end{figure}

For the virtual corrections to the process $\qbar q \rarr \tbar t$ we use the color decomposition
\beq
\label{eqn:col1Lqqb}
  A^\virt(\qbar q \rarr t\tbar) &=& \gs^4 \Big[
  \delta^{\bar{i}_1}_{i_4} \delta^{\bar{i}_3}_{i_2} \calB_1
  -\frac{1}{\Nc} \delta^{\bar{i}_1}_{i_2} \delta^{\bar{i}_3}_{i_4} \calB_2
  \Big].
\eeq
As the second term in \refeq{eqn:col1Lqqb} vanishes after interference with the leading order amplitude we do not consider it here.
The first term in \refeq{eqn:col1Lqqb} is given by
\beq
  \calB_1 &=& \Big(\Nc-\frac{2}{\Nc} \Big) \calA^a(1_{\tbar},2_t,3_{\qbar},4_q) - \frac{2}{\Nc} \calA^a(1_{\tbar},2_t,4_q,3_{\qbar})
  \nl
         &&  +\frac{1}{\Nc} \Big( \calA^b(1_{\tbar},2_t,3_{\qbar},4_q) + \calA^c(1_{\tbar},2_t,3_{\qbar},4_q) \Big)
	   - \sum_{f=1}^{\Nf} \calA^d_f(1_{\tbar},2_t,3_{\qbar},4_q).
\eeq
The ordered primitive amplitudes $\calA^{a,b,c,d}$ correspond to different topologies depending on which fermion lines enter the loop, cf. Fig.5.
\\

\section{}
\label{appB}

Following the notation of Ref.\cite{Catani:1996vz}, the generic form of
the subtraction terms for the partonic channels
$g(1)\, g(2) \rarr \tbar(3)\, t(4)\, g(5)$
is given by  twelve dipoles
\beq
   \calD^{15}_{3},\;
   \calD^{15}_{4},\;
   \calD^{25}_{3},\;
   \calD^{25}_{4},\;
   \calD^{15,2}_{},\;
   \calD^{25,1}_{},\;
   \calD^{1}_{35},\;
   \calD^{2}_{35},\;
   \calD^{1}_{45},\;
   \calD^{2}_{45},\;
   \calD^{}_{35,4},\;
   \calD^{}_{45,3}.
\label{eq15}
\eeq
It is easy to see that identical list of dipoles
is required to construct the subtraction terms for
$\qbar(1)\, q(2) \rarr \tbar(3)\, t(4)\, g(5)$ partonic channel. Note, however,
that the above dipoles differ for different partonic channels
since they contain particular  splitting kernels and reduced tree amplitudes.
The processes $\qparbar \! (1)\, g (2) \rarr \tbar (3)\, t(4)\, \qparbar \! (5)$ develop only collinear singularities making a summation over all spectator particles
unnecessary.
We therefore choose
\beq
\label{eqn:qgdipoles}
   \tilde{\calD}^{15,2}_{},\;
   \tilde{\calD}^{25,1}_{}
\eeq
as subtraction terms which differ from the original dipoles
Eq.(\ref{eq15}) only by their color factor.
In fact, since collinear singularities are local in color space,
no color correlations are required for the dipoles
in \refeq{eqn:qgdipoles}; the corresponding color factors are given by
$C_\mathrm{F}$ and $T_\mathrm{R}$ for each dipole in \refeq{eqn:qgdipoles}, respectively.

The auxiliary contribution from the dipole subtraction terms needs to be added back.
We choose to implement those terms by separately
integrating each dipole over the unresolved phase-space.
Singularities of  integrated dipoles  cancel against singularities  of  virtual corrections and unrenormalized parton distribution functions, yielding finite result for the cross-section. For the implementation of both
unintegrated and integrated dipoles we used results in Refs.~\cite{Catani:2002hc, Campbell:2004ch}.
Following Refs. \cite{Nagy:1998bb,Nagy:2003tz,Campbell:2004ch} we introduce a parameter $\alpha$ for all initial-state emitter dipoles which allows to
cut off the dipole phase-space in non-singular regions.
This helps to improve numerical stability of our code and provides a non-trivial check of the implementation of subtraction terms.

\end{appendix}

\end{document}